\documentstyle{article}

\begin{document}

\title{Did 20th century physics have the means to reveal the nature of inertia and
gravitation?}
\author{Vesselin Petkov \\
Physics Department, Concordia University\\
1455 de Maisonneuve Boulevard West\\
Montreal, Quebec H3G 1M8\\
vpetkov@alcor.concordia.ca \\
(or vpetkov@sympatico.ca)}
\date{14 December 2000}
\maketitle

\begin{abstract}
At the beginning of the 20th century the classical electron theory (or,
perhaps more appropriately, the classical electromagnetic mass theory) - the
first physical theory that dared ask the question of what inertia and mass
were - was gaining momentum and there were hopes that physics would be
finally able to explain their origin. It is argued in this paper that if
that promising research path had not been inexplicably abandoned after the
advent of relativity and quantum mechanics, the contemporary physics would
have revealed not only the nature of inertia, mass, and gravitation, but
most importantly would have outlined the ways of their manipulation. Another
goal of the paper is to try to stimulate the search for the mechanism
responsible for inertia and gravitation by outlining a research direction,
which demonstrates that the classical electromagnetic mass theory in
conjunction with the principle of equivalence offers such a mechanism.

\bigskip

\noindent NOTE: This paper presents only conceptual discussions of
rigorously obtained results which are available at:

\noindent http://alcor.concordia.ca/\symbol{126}vpetkov/papers/
\end{abstract}


\section{Introduction}

According to an eastern proverb the darkest place is beneath the lantern.
The meaning of a proverb could hardly be more profound and more suitable for
fundamental concepts: Nature has given us her greatest secrets as
self-evident phenomena. Everyone ''knows'' what existence, space, time,
mass, inertia, gravitation, etc. are; we even have complex theories dealing
with those concepts. If, however, we want to explain their nature we find
ourselves in the same situation in which St. Augustine found himself when
tried to explain the nature of time: ''What then is time? If no one asks me,
I know; if I wish to explain it to one that asketh, I know not'' \cite
{augustine}.

We are about to enter the $21$st century but our understanding of the origin
of inertia, mass, and gravitation still remains what has been for centuries
- an outstanding puzzle. All theories constituting the contemporary physics
are concerned mostly (if not only) with the description of those phenomena.
Even the modern theory of gravitation, general relativity, which provides a
consistent no-force explanation of gravitational interaction of bodies
following geodesic paths, is helplessly silent on the nature of the very
force we identify as gravitational - the force acting upon a body deviated
from its geodesic path being at rest on the Earth's surface.

The classical electromagnetic mass theory appears to have been gradually
forgotten and now many physicists believe that physics cannot say anything
about the origin of inertia, mass, or gravitation. Even scientists directly
involved in the efforts to discover the Higgs boson \cite{cern} (believed to
be responsible for endowing particles with mass) such as Claude Detraz, one
of the two research directors at $CERN$, think that ''Mass is a very
important property of matter, and we have nothing in our current theory that
says even a word about it''. In what follows we will see whether this is
really the case.

Here we shall follow the tradition established by the classical electron
theory (i.e. the classical electromagnetic mass theory) and will study the
inertial and gravitational properties of the simplest charged particle - the
classical electron. There are two reasons why the classical electron is
studied:

(i) There is no quantum mechanical model of the electron - quantum mechanics
describes only its state not the electron itself (later we will make an
attempt to outline the basis of the quantum electrodynamical formulation of
the electromagnetic mass theory), and

(ii) It is quite natural to complete the classical electromagnetic mass
theory first before making the transition to a quantum description of the
electron inertial properties. This has never been done since the classical
electromagnetic mass theory was virtually abandoned when the theory of
relativity and quantum mechanics were formulated. To abandon a promising
theory that has never been proven wrong is an unprecedented case in physics.
This unforgivable neglect is truly beyond one's comprehension because the
electromagnetic mass theory is even now the only theory that addresses the
origin of inertia and inertial mass in accordance with the experimental
evidence of the existence of electromagnetic inertia and of the
electromagnetic origin of some of the mass of charged particles. Moreover,
the classical electromagnetic mass theory predicted that the
(electromagnetic) mass increases with the increase of velocity (yielding the
correct expression) and that the relationship between energy and mass is $%
E=mc^{2}$ (see \cite{feynman}) - all this before the theory of
relativity.

\section{Classical electromagnetic mass theory}

In 1881 Thomson \cite{thomson} first realized that a charged particle was
more resistant to being accelerated than an otherwise identical neutral
particle and conjectured that inertia can be reduced to electromagnetism.
Due mostly to the works of Heaviside \cite{heaviside}, Searle \cite{searle},
Lorentz \cite{lorentz}, Poincar\'{e} \cite{poincare}, Abraham \cite{abraham}%
, Fermi \cite{f=ma}, Mandel \cite{mandel}, Wilson \cite{wilson},
Pryce \cite {pryce}, Kwal \cite{kwal}, and Rohrlich
\cite{rohrlich} this conjecture was developed into a theory (the
classical electromagnetic mass theory of the electron) in which
inertia is a local phenomenon originating from the interaction of
the electron charge with itself (i.e. with its own electromagnetic
field) \cite{history}.

According to the classical model of the electron its charge is uniformly
distributed on a spherical shell. Such a model, however, cannot explain why
the electron is stable since the negatively charged spherical shell tends to
blow up due to the mutual repulsion of the different ''parts'' of the
charge. This difficulty, known as the stability problem of the electron, has
two sides - computational and conceptual. In the beginning of the century it
appeared that the stability problem did lead to a computational difficulties
with the famous $4/3$ factor stubbornly appearing in the different
expressions for the electromagnetic mass. Several authors \cite{fermi}-\cite
{rohrlich} independently showed that the $4/3$ factor had been caused by
incorrect calculations, not by the model itself. This implies that there is
no real problem with the stability of the electron. We do not know why. What
we do know, however, is that if there were such a problem it would
inevitably show up in all calculations of the electromagnetic mass which is
not the case.

The conceptual difficulty with the classical model of the
electron - its failure to explain what prevents the electron
charge from blowing apart - has never been satisfactory resolved
by the classical electromagnetic mass theory \cite{poincare2}.
This is viewed as an indication that that model is not entirely
correct. There are three reasons, however, which demonstrate that
the classical model of the electron works. (i) The very existence
of the radiation reaction force is evidence that there is
interaction (repulsion) between the different parts of the
electron charge. "The radiation reaction is due to the force of
the charge on itself - or, more elaborately, the net force
exerted by the fields generated by different parts of the charge
distribution acting on one another" \cite{griffiths} (in the case
of a single radiating electron the presence of a radiation
reaction implies interaction of different parts of the electron).
This may indicate that the charge of the real electron can indeed
be modeled by a small spherical shell which, but the electron
charge should be not {\em continuously} existing as smeared out
on the shell; only if the electron charge were occupying the whole
shell at every instant the stability problem would arise
\cite{elementary}. (ii) The calculations of the electron
electromagnetic mass (assuming a spherical distribution of its
charge) yield the correct expression for the mass. (iii) A more
important indication that the classical model of the electron
cannot be discarded as inadequate is that the classical
electromagnetic theory is the {\em only} theory that correctly
predicts the experimental fact that at least part of the
electron's inertia and mass are electromagnetic in origin. As
Feynman put it: "There is definite experimental evidence of the
existence of electromagnetic inertia - there is evidence that
some of the mass of charged particles is electromagnetic in
origin" \cite[p. 28-10]{feynman}, \cite{feynman1}.

At the beginning of the century many physicists recognized ''the tremendous
importance, which the concept of electromagnetic mass possesses for all of
physics: It is the basis of the electromagnetic theory of matter'' (E. Fermi
\cite{moylan}). Therefore, it would have been natural to develop further the
theory of electromagnetic mass by taking into account the relevant new
results in physics achieved in this century. Instead, it had been
inexplicably abandoned: ''The state of the classical electron theory reminds
one of a house under construction that was abandoned by its workmen upon
receiving news of an approaching plague. The plague in this case, of course,
was quantum theory. As a result, classical electron theory stands with many
interesting unsolved or partially solved problems'' (P. Pearle \cite{pearle}%
).

It is clear that what Fermi, Feynman, and Pearle are saying is important but
not crucial (one can always find favourable quotes). What is crucial is what
the classical electromagnetic mass theory itself is saying on inertia, mass,
and gravitation when the principle of equivalence is taken into account.

The mechanism responsible for the electron's inertia and mass according to
the classical electromagnetic mass theory is the following. The repulsion of
the charge elements of an electron in uniform motion in flat spacetime
cancels out exactly and there is no net force acting on the electron. If,
however, the electron is accelerated the repulsion of its volume elements
becomes unbalanced and as a result it experiences a net self-force ${\bf F}%
_{self}$ which resists its acceleration - it is precisely this resistance
that we call inertia (for a detailed description of why the repulsion of the
different parts of an accelerating electron becomes unbalanced see \cite[p.
28-5]{feynman}). The self-force is opposing the external force that
accelerates the electron (i.e. its direction is opposite to the electron's
acceleration ${\bf a}$) and turns out to be proportional to ${\bf a}$: ${\bf F}_{self}$ $=-m%
{\bf a}$, where the coefficient of proportionality $m$ represents
the inertial mass of the electron and is equal to $E/c^{2}$ where
$E$ is the energy of the electron field (therefore the electron
inertial mass is electromagnetic in origin). This is an amazing
result for three reasons: (i) it reveals that both inertia and mass have electromagnetic origin; the mass $%
m$ in the expression for the self-force is electromagnetic since it is
simply the mass that corresponds to the energy of the electron's electric
field through the relation $E=mc^{2}$; (ii) it demonstrates that inertia is
a local phenomenon (contrary to Mach's hypothesis that the local property of
inertia has a non-local origin \cite{mach}) and (iii) it constitutes the
first derivation of Newton's second law ${\bf F}=m{\bf a}$ \cite{elementary}
- a law that is considered so fundamental that after Newton postulated it no
one has attempted to derive it.

Therefore, the classical electromagnetic mass theory does say not only a
word, but offers a detailed mechanism explaining the origin of inertia and
mass of charged particles: it is the {\em unbalanced} repulsion of the
volume elements of the charge of an accelerating electron that gives rise to
the electron's inertia and inertial mass.

An observer at rest in an accelerating reference frame will see
that the electric field of an electron also at rest in the frame
is distorted. Unlike uniform velocity, acceleration is absolute
and the distorted electric field of the electron is one of the
means by which the observer can detect the frame's acceleration.
Therefore, in terms of the distorted electric field of an
accelerating electron and avoiding the use of the controversial
concept "parts of an elementary charge", one can equivalently say
that an electron's inertia and inertial mass originate from the
interaction of its charge with its own distorted electric field.
The interaction of the charge of an uniformly moving electron
with its own Coulomb (not distorted) field produces no net force
acting on the electron as a whole; that is why an electron moving
with constant velocity offers no resistance to its uniform motion
\cite{repulsion}.

The electromagnetic mass theory has been not only gradually
forgotten; its status is now even more awkward - those who
mention it regard the electron mass as electromagnetic only in
part as if the $4/3$ factor in the expression for the
electromagnetic mass has not been accounted for. It is that factor
that was considered an indication that not the entire electron
mass was electromagnetic. Feynman's objection against regarding
the entire electron mass as electromagnetic was that the $4/3$
factor leads to a contradiction with special relativity \cite[p.
28-4]{feynman}. Now, after the removal of that factor, it clearly
follows from the classical electromagnetic mass theory that the
entire mass of the electron should be electromagnetic in origin
\cite{phd}.

\section{Electromagnetic mass theory and the principle of equivalence}

The classical electromagnetic mass theory offered a mechanism
accounting for the origin of inertia and inertial mass, but
before the formulation of the equivalence principle by Einstein
it appeared that that theory does not explain the origin of the
passive gravitational mass and does not affect gravitation at
all. The equivalence principle, however, postulated that the
inertial mass (the measure of resistance that a body offers when
accelerated) is equal to the (passive) gravitational mass (the
measure of resistance that a body offers when being prevented
from falling in a gravitational field); L. von E\"{o}tv\"{o}s'
experiments had already confirmed that equality. The equivalence
principle requires that inertial and gravitational mass be equal
but it was not initially clear how the electromagnetic mass
theory could explain the origin of the gravitational mass in a
gravitational field. The answer to this question is that it is a
spacetime anisotropy around massive bodies that is responsible
for the force acting upon an electron on the Earth's surface and
for its gravitational mass. It manifests itself in the anisotropy
in the velocity of electromagnetic signals (for short - the
velocity of light). To explain what is the origin of the passive
gravitational mass according to the electromagnetic mass theory
and to shed some light into the basis of the equivalence
principle here is a brief description of what happens to an
electron in an accelerated reference frame and a frame of
reference supported in a gravitational field.

\subsection{An electron in an accelerated reference frame $N^{a}$}

For an observer at rest in an inertial reference frame $I$ the
electromagnetic field of an accelerating electron is distorted
due to the electron's accelerated motion. As the accelerated
motion is absolute the electron's electric field will be also
distorted for an observer at rest in the accelerating
(non-inertial) reference frame $N^{a}$ in which the electron is
at rest. The distortion of the electron's field for the inertial
observer in $I$ is caused by the electron's accelerated motion.
For the non-inertial observer in $N^{a}$, however, the electron
is at rest and therefore there is no (accelerated) motion of the
electron that can account for the distortion of its field as
determined by the observer in $N^{a}$. What causes the
deformation of the electron's field in $N^{a}$ is the anisotropic
velocity of light there; $N^{a}$ is an accelerating frame and it
is the anisotropy in the propagation of light (and its
manifestations such as the distorted
electron field) which allow an observer in $N^{a}$ to determine from within $%
N^{a}$ that it is an accelerating (non-inertial) frame (for a more detailed
discussion why the velocity of light in a non-inertial frame is anisotropic
see \cite{phd} and \cite{petkov}).

\subsubsection{An electron falling in $N^{a}$}

Imagine that an inertial observer $I$ is observing an electron
floating inside a spacecraft which moves with a constant velocity
with respect to $I$ (so both the spacecraft and the electron move
by inertia offering no resistance to their motion). Let's now
assume that the spacecraft starts to accelerate with an
acceleration ${\bf a}$, i.e. it becomes a non-inertial frame
$N^{a}$ (an observer in the spacecraft will be also called
$N^{a}$). For the inertial observer $I$ nothing happens to the
electron - it continues to move by inertia until the spacecraft's
floor reaches it. For an observer in the spacecraft, however, the
electron is falling toward the floor with an acceleration ${\bf
a}$ (for $I$ it is the spacecraft's floor that approaches the
electron). Obviously, there is a problem here - as the {\em
non-resistant} motion by inertia is absolute, both observers ($I$
and $N^{a}$) should agree that the electron is moving by inertia
inside the spacecraft which does not appear to be the case since
for $N^{a}$ the electron is accelerating toward the floor (which
implies that there is a force that accelerates it). Despite the
fact that today's physics regards the force accelerating the
electron inas fictitious, it has never explained why the {\em
accelerated} motion of the electron as viewed in $N^{a}$ should be
considered {\em inertial} in $N^{a}$ as well.

When the anisotropic velocity of light in $N^{a}$ is taken into account in
the calculation of the electric field of the falling in $N^{a}$ electron it
turns out that at every instant the electron field is the Coulomb (not
distorted) field (here the instantaneous field is considered in order to
separate the Lorentz contraction of the field and the distortion due to
acceleration) \cite{phd}. Therefore, for an observer in $N^{a}$ the motion
of the falling (accelerating) electron will not be resistant since its
electric field is not distorted. This means that the free electron in $N^{a}$
falls with an acceleration ${\bf a}$ in order to compensate the anisotropy
in the propagation of light in $N^{a}$ and to prevent its field from being
distorted; in other words, the falling electron offers no resistance to its
accelerated motion in $N^{a}$ and therefore moves by inertia while falling
in $N^{a}$. In such a way, as expected, both the inertial observer $I$ and
the non-inertial observer $N^{a}$ agree that the electron in $N^{a}$ is
moving by inertia offering no resistance to its motion.

\subsubsection{An electron at rest in $N^{a}$}

Now consider the moment when the spacecraft's floor reaches the floating
electron as seen by the inertial observer $I$. The electron starts to
accelerate and its motion is no longer non-resistant; its field gets
distorted and a self-force ${\bf F}_{self}$ $=-m{\bf a}$ originating from
the unbalanced repulsion of the electron's charge ''elements'' (caused by
the accelerated motion of the electron) starts to oppose its acceleration
(i.e. the deformation of its field); here again $m$ is the electromagnetic
mass of the electron which is the mass corresponding to the energy of the
electron field $E$ ($m=E/c^{2}$).

What can an observer in $N^{a}$ (in the spacecraft) say about the electron
on the spacecraft's floor? At first, it appears that the electron field is
not distorted with respect to $N^{a}$ since it is at rest in $N^{a}$ which
would mean that no force is acting on the electron. If this were the case,
there would be a problem again: the inertial and the non-inertial observers
would differ on whether the electron is subjected to a force; as the
existence of a force is an absolute fact all observers should recognize it.
That problem disappears when the anisotropic velocity of light is taken into
account in the calculation of the electron field in $N^{a}$. Due to an
unnoticed up to now Li\'{e}nard-Wiechert-like contribution to the potential
of a charge in a non-inertial reference frame \cite{phd}, \cite{petkov}
(caused by the anisotropic velocity of light in such frames) the electric
field of the electron in $N^{a}$ is as distorted as the field seen by the
inertial observer $I$. Therefore, the non-inertial observer $N^{a}$ will
also find that the electron is subjected to the purely electric self-force $%
{\bf F}_{self}$ $=-m{\bf a}$, originating from the anisotropic velocity of
light in $N^{a}$ which disturbs the balance of the mutual repulsion of the
''elements'' of the electron charge. As seen from the expression for the
self-force it coincides with what we call the inertial force; hence it
follows that the inertial force is electromagnetic in origin.

The non-inertial observer in $N^{a}$ sees that when the falling electron
reaches the floor of the spacecraft it can no longer compensate the
anisotropy in the propagation of light in $N^{a}$ (by falling with an
acceleration ${\bf a}$), its field gets distorted which gives rise to the
self-force ${\bf F}_{self}$.

\subsection{An electron in a gravitational field}

Consider now an electron at rest in the Earth's gravitational field. The
Newtonian theory of gravitation tells us that the electron is subjected to a
gravitational force - its weight ${\bf F}=m{\bf g}$. What does general
relativity say about that force? Nothing. The gravitational field in general
relativity is a manifestation of spacetime curvature and (unlike the
electromagnetic field) is not a force field (which means that ''there is no
gravitational force in general relativity'' \cite{synge}). A body falling
toward the Earth is represented by a geodesic worldline which means that no
force is acting on it. If a body is on the Earth's surface, however, its
worldline is no longer geodesic and it is subjected to a force whose nature
is an open question in general relativity \cite{petkov}. This fact alone
(not to mention the issue of the represented by a {\em pseudo}-tensor energy
and momentum of the gravitational field \cite{graven}) is a sufficient
reason for a thorough re-examination of the foundations of general
relativity. And this is urgently needed since, as we shall see bellow, there
are strong arguments indicating that the correct interpretation of the
formalism of general relativity should be in terms of anisotropic, not
curved spacetime.

\subsubsection{An electron at rest in the Earth's gravitational field}

One of the formulations of the equivalence principle states that what is
happening in a non-inertial reference frame $N^{a}$ which accelerates with
an acceleration ${\bf a}$ also happens in a non-inertial reference frame $%
N^{g}$ at rest in a gravitational field characterized by an acceleration $%
{\bf g}=-{\bf a}$. One of the results Einstein obtained by analyzing the
principle of equivalence is that in two elevators - one accelerating (frame $%
N^{a}$) and another at rest in the Earth's gravitational field (frame $N^{g}$%
) - light bends when propagating perpendicularly to the accelerations ${\bf a%
}$ and ${\bf g}$, respectively. If one considers light
propagating parallel and anti-parallel to ${\bf a}$ and ${\bf
g}$, it turns out that the average velocity of light in $N^{a}$
and $N^{g}$ is anisotropic: the velocity of a light ray from the
elevator's ceiling toward the floor is slightly greater that the
velocity of light propagating in the opposite direction
\cite{phd}, \cite{petkov}. Interestingly, the expression for the
average anisotropic velocity of light follows from the expression
of the velocity of light in a gravitational field obtained by
Einstein in 1911 but abandoned when the calculations of the
deflection of light by the Sun (based on that expression)
predicted a wrong value for the deflection angle. A careful
analysis of the propagation of light in the Einstein thought
experiment involving the two elevators demonstrates that his 1911
expression for the velocity of light in a gravitational field has
been prematurely discarded \cite{phd}.

Due to the anisotropic velocity of light in $N^{g}$ the electric field of an
electron at rest in $N^{g}$ distorts, the balance of the mutual repulsion of
the electron charge ''elements'' is disturbed which in turn gives rise to a
self-force ${\bf F}_{self}$ which tries to restore the balance in the mutual
repulsion. The self-force turns out to be ${\bf F}_{self}$ $=m{\bf g}$,
where $m=E/c^{2}$ represents the passive gravitational mass of the electron
and $E$ is the energy of its field. As the electric self-force ${\bf F}%
_{self}$ is precisely equal to the gravitational force ${\bf F}=m{\bf g}$,
the classical electromagnetic mass theory predicts that the gravitational
force acting on an electron on the Earth's surface is purely electromagnetic
in origin which means that its passive gravitational mass is also
electromagnetic in origin.

This is an important result since it demonstrates that the
self-forces ${\bf F}_{self}$ $=-m{\bf a}$ in $N^{a}$ and ${\bf F}_{self}$ $=m{\bf g}$ in $%
N^{g}$ have precisely the {\em same} origin: in both cases it is the
anisotropic velocity of light (electromagnetic signals) that gives rise to
the electric force ${\bf F}_{self}$ by distorting the electric field of the
electron at rest in $N^{a}$ and $N^{g}$ which in turn disturbs the balance
in the repulsion of its charge ''elements''. What we call the inertial mass $%
m$ (in ${\bf F}_{self}$ $=-m{\bf a}$) and the passive gravitational mass $m$
(in ${\bf F}_{self}$ $=m{\bf g}$) are precisely the {\em same} thing: $m$ is
the measure of the resistance an electron offers when its field is being
distorted. In the case of an accelerating electron it is its acceleration
(i) that distorts its field as seen by an inertial observer and (ii) that
causes the anisotropy in the propagation of light in $N^{a}$ which in turn
distorts the electron field as observed by a non-inertial observer in $N^{a}$%
. Similarly, the distortion of the field of an electron at rest on the
Earth's surface (i.e. at rest in $N^{g}$) is caused by the anisotropic
velocity of light in $N^{g}$.

\subsubsection{An electron falling in the Earth's gravitational field}

The self-force ${\bf F}_{self}$ acting on an electron at rest on the Earth's
surface arises on account of its distorted electric field (caused by the
anisotropic velocity of light in $N^{g}$) which disturbs the balance in the
mutual repulsion of the electron charge ''elements''. ${\bf F}_{self}$ tries
to prevent the electron field from distorting and to restore the repulsion
balance. If we allow ${\bf F}_{self}$ do its job by removing the obstacle
beneath the electron, it will start to fall and it will fall in such a way
that the distortion of its field is eliminated, the repulsion balance is
restored and the self-force ${\bf F}_{self}$ ceases to exist. The
calculation of the electric field of an electron left on itself in a
gravitational field shows that the only way for the electron to compensate
the anisotropy in the propagation of light in the gravitational field and to
prevent its field from being distorted is to fall with an acceleration ${\bf %
g}$ \cite{phd}. Therefore, a free electron in a gravitational field will
move by inertia (without resistance) only if it falls with an acceleration $%
{\bf g}$. This result sheds light on the fact that in general relativity the
motion of a body falling toward a gravitating center is regarded as inertial
(non-resistant) and is represented by a geodesic worldline. Therefore, the
electromagnetic mass theory gives an elegant answer to the question why an
electron is falling in a gravitational field and no force is causing its
acceleration.

The result that the electric field of an electron falling in the Earth's
gravitational field at any instant is the Coulomb field, which means that no
self-force is acting on the electron, also demonstrates that a falling
electron does not radiate - its electric field is the Coulomb field and
therefore does not contain the radiation $r^{-1}$ terms \cite{phd}.

If the electron is prevented from falling its electric field
distorts, the self-force ${\bf F}_{self}$ appears and tries to
force the electron to move (fall) in such a way that its field
becomes the Coulomb field; as a result of the free fall of the
electron the self-force disappears.

\subsubsection{The electromagnetic mass theory explains the behaviour of an electron in a gravitational field}

The behaviour of the classical electron in a gravitational field is {\em %
fully} accounted for by the classical electromagnetic mass theory and the
equivalence principle: the anisotropic velocity of light in $N^{g}$ (in an
elevator at rest in the Earth's gravitational field)

(i) gives rise to a self-force acting on an electron at rest in
$N^{g}$ (whose worldline is deviated from its geodesic status) by
distorting the electric field of the electron which in turn
disturbs the balance in the mutual repulsion of its charge
''elements'', and

(ii) makes a free electron fall in $N^{g}$ with an acceleration ${\bf g}$ in
order to preserve its Coulomb field and therefore to balance the repulsion
of its charge ''elements''. No force is acting upon a falling electron
(whose worldline is geodesic) but if it is prevented from falling (i.e.
deviated from its geodesic path) the mutual repulsion of the ''elements'' of
its charge becomes unbalanced which results in a self-force trying to force
the electron to fall.

General relativity does not provide an explanation of the nature of the
force acting on a body at rest in a gravitational field whose worldline is
not geodesic. It appears that general relativity cannot provide such an
explanation at all since ''there is no gravitational {\em force} in general
relativity'' \cite{synge}; this fact constitutes not only an open question
but a crisis in general relativity. The classical electromagnetic mass
theory in conjunction with the principle of equivalence provides a natural
answer to the questions (i) why a free electron in a gravitational field is
falling by itself (with no force acting upon it) and why its worldline is
geodesic, and (ii) why an electron at rest in a gravitational field is
subjected to a force and why its worldline is not geodesic: the worldline of
an electron which preserves the shape of its Coulomb field is geodesic and
represents a free non-resistantly moving electron; if the field of an
electron is distorted, its worldline is not geodesic and the electron is
subjected to a self-force on account of its own distorted field.

The anisotropic velocity of light in both $N^{a}$ and $N^{g}$ is responsible
for the fall of a free electron and the appearance of a self-force when the
electron is prevented from falling in $N^{a}$ and $N^{g}$. Therefore, it is
the anisotropy in the propagation of light in $N^{a}$ and $N^{g}$ that makes
the two non-inertial reference frames $N^{a}$ and $N^{g}$ equivalent. In
such a way, the equivalence principle is a straightforward corollary of the
anisotropic propagation of light in $N^{a}$ and $N^{g}$. The anisotropy in
the velocity of light in the accelerating reference frame $N^{a}$ is caused
by the frame's acceleration. What is the origin of the anisotropic velocity
of light in the non-inertial reference $N^{g}$ (at rest on the Earth's
surface) will be discussed bellow.

\subsection{Toward a general theory of motion and gravitation?}

In a spacetime region where the propagation of light is isotropic a free
electron does not resist its motion only if it moves with uniform velocity
(which means that its electric field is not distorted - it is the Coulomb
field \cite{distortion}); in this case the electron's worldline is a
straight geodesic line. If the electron is prevented from moving with
constant speed its field distorts and the electron resists its acceleration
(i.e. it resists the distortion of its field); in this case the worldline of
the accelerating electron is neither geodesic nor straight. In a spacetime
region where the propagation of light is anisotropic (i.e. in an elevator on
the Earth's surface) the motion of a free electron is non-resistant
(preserving the Coulomb shape of its field) only if it falls with an
acceleration ${\bf g}$; in this case the electron's worldline is geodesic
but not a straight line. If the electron is prevented from falling (i.e.
from moving by inertia in an anisotropic region of spacetime) its field
distorts and the electron resists the deformation of its field; in this case
its worldline is neither geodesic nor straight.

The electromagnetic mass theory provides an amazingly elegant and
consistent description of the motion of the classical electron in
open space (far from massive objects) and in a gravitational
field. If it turns out that a quantum electrodynamical
formulation of the electromagnetic mass theory reproduces the
same elegant picture of the motion of charged particles we will
realize that the basis for a general theory of motion and
gravitation has been available for a century now.

\section{Spacetime curvature or spacetime anisotropy?}

We have seen that the anisotropy in the propagation of light and
the electromagnetic mass theory fully account for the behaviour of
the classical electron in a gravitational field. It appears that
no curvature of spacetime is needed. In order to see whether this
is really the case let us first consider what causes the
gravitational attraction of two electrons.

We have seen that the electron inertial and passive gravitational
mass are entirely electromagnetic in origin. As it is believed
that all three masses - inertial, passive gravitational, and
active gravitational - are equal, it follows that the electron
active gravitational mass is fully electromagnetic in origin as
well. And since it is only the charge of the electron that
represents it (there is no mechanical mass), it follows that the
active gravitational mass of the electron is represented by its
charge. Therefore it is the electron charge that causes its
gravity and the anisotropic velocity of light in the electron's
neighborhood. The question now is: ''Is the electron's
gravitational field a manifestation of a spacetime curvature
around the electron?'' or more precisely: ''Does the electron's
charge create a curvature which in turn causes the anisotropic
velocity of light in the electron's vicinity?''

We have seen that it is the anisotropy in the velocity of light
and the electromagnetic mass theory that fully and consistently
explain the fall of an electron toward the Earth and the
self-force acting on an electron at rest on the Earth's surface.
Let us now see whether the (gravitational) attraction between two
electrons can be explained in the same way.

In addition to the electric repulsion of two electrons ($e_{1}$
and $e_{2}$) in open space, they also attract each other through
the anisotropy in the velocity of light around each of them:
$e_{1}$ falls toward $e_{2}$ in order to compensate the
anisotropy caused by $e_{2}$ and vice versa. Therefore, the
anisotropy in the propagation of light in the electrons' vicinity
and the electromagnetic mass theory are completely sufficient to
explain the electrons' (gravitational) attraction. No additional
spacetime curvature hypothesis is necessary. This is an
indication that, according to the classical electromagnetic mass
theory, what the electron's charge creates is not a spacetime
curvature; it is a spacetime anisotropy which causes the
anisotropic velocity of light. In such a way, an electron's
gravitational field according to the electromagnetic mass theory
turns out to be the spacetime anisotropy in the neighborhood of
the electron originating from its charge.

An anisotropic-spacetime re-interpretation of general relativity
appears more trouble-free than its current curved-spacetime
interpretation: (i) there is no problem with the force acting on a
body whose worldline is not geodesic; (ii) the problem with the
existence of gravitational energy is solved (there is energy, but
electromagnetic); (iii) the equivalence of an accelerating frame
and a frame at rest in a gravitational field is explained along
with the equivalence of inertial and passive gravitational mass;
(iv) it is also explained what tells a body to move
non-resistantly (by inertia) or to offer resistance: a free body
moves by inertia if the electric field of each of its charges is
the Coulomb field; if the body's charges' fields are distorted,
the body's motion is with resistance - it resists the deformation
of its charges' electric fields.

\section{Toward a quantum electrodynamical formulation of the
electromagnetic mass theory}

Although the lack of a quantum model of the electron makes it impossible to
formulate the classical electromagnetic mass theory in terms of quantum
mechanics, it appears that its quantum electrodynamical formulation may be
possible.

According to the electromagnetic mass theory the inertial and
gravitational force acting on the classical electron originate
from its self-interaction through its distorted field. In quantum
electrodynamics (QED) the quantized electric field of a charge is
represented by a swarm of virtual photons that are constantly
being emitted and absorbed by the charge. The distorted field of a
non-inertial electron in QED is represented by the anisotropy in
the velocity of the virtual photons comprising the electron
field. The recoil that an electron experiences every time it emits
or absorbs a virtual photon depends on the photon's velocity.
Therefore, the anisotropy in the virtual photons' velocity
disturbs the balance of the recoils to which a non-inertial
electron is subjected (all recoils cancel out exactly if there is
no anisotropy). This means that in QED too the interaction of a
non-inertial electron with its own distorted field also gives
rise to a self-force which may completely coincide with the
inertial force in the case of an accelerating electron and with
the gravitational force in the case of an electron supported in a
gravitational field.

It turns out that in the QED formulation of the electromagnetic
mass theory it does not matter whether an electron will be
regarded as a point or a sphere - in both cases the self-force
acting on a non-inertial electron originates from the {\em
unbalanced} recoils of the virtual photons being absorbed by the
electron; the recoils of the emitted virtual photons cancel out
since the photons are always emitted with initial speed $=c$.

The QED formulation of the electromagnetic mass theory makes it
possible to compare the electromagnetic mass approach to inertia
with a recently proposed zero-point field (ZPF) approach to
inertia \cite{hrp} according to which the inertia of an
(accelerating) electron originates from the interaction of its
charge with the virtual photons of the zero-point fluctuation of
the electromagnetic vacuum. A careful analysis of the two
approaches demonstrates that they are regarding different
mechanisms as the origin of inertia: the self-force acting on a
non-inertial charge (according to the electromagnetic mass
theory) originates from the disturbed balance of the recoils from
the virtual photons comprising the charge's own electric field
while the reaction force an accelerating charge is experiencing
(according to the ZPF approach) is caused by the ZPF fluctuations
of the entire electromagnetic vacuum. There are two reasons which
seem to indicate that the ZPF inertia may be only a small
contribution to the electron's inertia caused by the unbalanced
recoils from the virtual photons of its field:

(i) The ZPF approach appears unable to explain gravitation and the
origin of gravitational mass. The electromagnetic resistance
offered by an accelerating charge as explained by the ZPF
approach can be viewed as caused by the Lorentz force which
originates from the interaction of the {\em magnetic} component of
the electromagnetic ZPF with the charge. This addresses the
charge inertial mass only. When the charge is at rest on the
Earth's surface, however, there is no magnetic component of the
electromagnetic ZPF which can interact with the charge and
therefore there is no ZPF contribution to the charge
gravitational mass. One may speculate that inertial and
gravitational mass are not equal at the quantum level which is,
of course, something that should be studied. It becomes clear
from here, however, why the ZPF contribution to inertia will be a
small correction at best if there is no ZPF contribution to the
gravitational mass: a ZPF contribution to the inertial mass
should be extremely small since a greater contribution will
result in a greater difference between the two masses which would
have been observed by now.

(ii) As the classical electromagnetic mass theory is the only
classical theory (supported by experimental evidence) that deals
with inertia it is natural to expect that it should be the QED
version of that theory that accounts for inertia in QED. On the
other hand, the ZPF approach to inertia does not have a classical
analog which seems to disqualify it as a theory describing the
major contribution to inertia.

As inertia and gravitation have predominantly macroscopic
manifestations it appears certain that these phenomena should
possess not only a quantum but a classical description as well.
This expectation is corroborated by the fact that such a
description already exists - the electromagnetic mass theory
which yields the correct expressions for the inertial and
gravitational mass of the classical electron. Therefore, the
chances of any modern theory of inertia (and gravitation) can be
evaluated by seeing whether it can be considered a quantum
generalization of the classical electromagnetic mass theory.

\section{Is all the mass electromagnetic?}

We have seen that both the inertial and the passive gravitational
mass of the classical electron are fully electromagnetic in
origin. If we now ask what about the inertial and gravitational
mass of the real electron? Are they electromagnetic in origin as
well?

An argument against regarding the entire mass as electromagnetic
is that strong and weak interactions should also contribute to
the mass. This argument, however, does not apply to the electron
for two reasons: (i) the electron does not participate in strong
interactions, and (ii) a free electron does not participate in
any weak interactions either.

This argument, however, is quite relevant when the nature of mass
of the other elementary charged particles is discussed. As the
issue of the strong and weak contributions to the mass is an open
one and needs a separate study, let us outline an argument
demonstrating that at least the strong interaction does not
contribute to the mass. As we have seen the unbalanced {\em
repulsion} of the charge ''elements'' of the classical electron
gives rise to its inertia and mass. Therefore the unbalanced {\em
repulsion} of two like charges increases the mass of the
two-charge system. Unbalanced {\em attraction} of opposite charges
results in the reduction of the charges' mass \cite{phd}. This is
true not only for electric forces. Early attempts by Poincar\'{e}
\cite {poincare} to resolve the stability problem in the classical
electromagnetic mass theory resulted in the introduction of
unknown attraction forces (called Poincar\'{e} stresses) that
balance the repulsion of the charge ''elements'' of the classical
electron. As it turned out that those
attraction forces had a negative contribution to the mass the problematic $%
4/3$ factor was reduced to $1$. Therefore, due to (i) the fact that the
forces of strong interaction are attraction forces and (ii) the strength of
strong interaction (over two orders of magnitude greater than the
electromagnetic interaction) one can expect a significant negative
contribution to the mass of a charged particle (compared to the
electromagnetic contribution). If it turns out that the strong interaction
does contribute to the mass, we will face a major crisis in physics - it
will not be clear what compensates the negative contribution to the mass
that originates from the strong interaction.

On the other hand, however, the strong and weak interactions as fundamental
forces should make a contribution to the mass (as the electromagnetic
interaction does) \cite{stephani} and if they do not, then we might be
forced to re-examine their very nature as separate fundamental interactions.

If it turns out that the strong and weak interactions make no contribution
to the mass then the mass of all particles will prove to be entirely
electromagnetic in origin. It should be noted, however, that a fully
electromagnetic mass implies that there are no elementary neutral particles
(with non-zero rest mass) in nature. A direct consequence from here is that
only charged particles or particles that consists of charged constituents
possess inertial and passive gravitational mass. Stated another way, it is
only elementary charges that comprise a body; there is no such fundamental
quantity as mass. It is evident that in this case the electromagnetic mass
theory predicts zero neutrino mass and appears to be in conflict with the
apparent mass of the $Z^{0}$ boson which is involved in the weak
interactions. The resolution of this apparent conflict could lead to either
restricting the electromagnetic mass theory (in a sense that not the entire
mass is electromagnetic) or re-examining the facts believed to prove (i)
that the $Z^{0}$ boson is a fundamentally neutral particle (unlike the
neutron), and (ii) that it does possess inertial and gravitational mass if
truly neutral.

Another argument that the mass of a particle is fully electromagnetic in
origin comes from the velocity dependence of the mass. It is a corollary of
the classical electromagnetic mass theory that the electromagnetic mass
rises with velocity inversely as $\left( 1-v^{2}/c^{2}\right) ^{1/2}$
\cite[p. 28-3]{feynman}. And instead of viewing the result that all the mass
depends on velocity discovered by the special theory of relativity as a
serious indication that all the mass is electromagnetic, inexplicably the
whole issue of electromagnetic mass has been practically abandoned. If we
assume that the mass of a body consists of several kinds of masses
(electromagnetic, mechanical, strong and weak) we have to answer the
question how all of them obey the {\em same} law of velocity dependence?

\section{Conclusions}

We have seen that inertia, inertial mass, gravitation, passive
gravitational mass, and the equivalence of the two masses of the
classical electron are fully accounted for by the electromagnetic
mass theory. The self-force to which a non-inertial electron is
subjected on account of its own distorted electric field
unambiguously indicates that the inertial and passive
gravitational mass of the classical electron are electromagnetic
in origin. This result provides a straightforward explanation of
the equivalence of inertial and gravitational mass. The inertial
and passive gravitational mass of the electron are the {\em same}
thing - the mass that corresponds to the energy of its electric
field. However, the inertial and passive gravitational mass of
the electron manifest themselves as such - as a measure of the
electron's resistance to being accelerated - only if it is
subjected to an acceleration (kinematic or gravitational). This
resistance originates from the unbalanced mutual repulsion of the
volume elements of the electron.

According to the electromagnetic mass theory it is the electron
charge that causes the anisotropy in the propagation of light in
the electron's vicinity. That anisotropy completely accounts for
the (gravitational) attraction of two electrons. This is an
indication that the Riemann curvature tensor in general relativity
should be regarded as describing an anisotropy (not curvature) of
spacetime. An important open question of the electromagnetic mass
theory is how a charge causes the anisotropy of spacetime around
itself. The study of this question may have important
implications for the possibility of controlling gravitational
interaction one day.

It should be stressed that the electromagnetic mass theory is not just a
hypothesis; it is a valid physical theory since (i) it is based on firm
experimental evidence (the experimental fact that at least part of the mass
of charged particles is electromagnetic in origin; there is no other theory
that accounts for this fact), and (ii) it is a further natural development
of the classical electron theory in conjunction with the principle of
equivalence.

It should be also specifically emphasized that even if it turns
out that only part of the mass is electromagnetic in origin it
still follows from the electromagnetic mass theory that inertia,
inertial mass, gravitation, and gravitational mass (passive and
active) are in part electromagnetic in origin.

The fact that a valid physical theory predicts that inertia and
gravitation are entirely or at least partly electromagnetic in
origin should have received the attention this incredible result
deserves. The prediction that it is electromagnetic phenomena that
cause (fully or partly) inertia and gravitation means that they
can be in principle manipulated since we know how to deal with
electromagnetic phenomena.

All results discussed here could have been obtained at least
eighty years ago when Fermi \cite{fermi} initiated the approach of
studying the classical electromagnetic mass theory in conjunction
with general relativity (more specifically, with the equivalence
principle). Unfortunately, he later turned to atomic physics and
perhaps deprived our century from solving the mystery of inertia
and gravitation.

I believe that the answer to the question posed in the title is now clear -
Yes, the $20th$ century physics did have the means to reveal the nature of
inertia and gravitation.


\begin{thebibliography}{99}

\bibitem{augustine}  St. Augustine, Confessions, Book $11(11.14.17)$;
http://ccat.sas.upenn.edu/jod/augustine.html (see the Section ''Texts and
Translations'').

\bibitem{cern}  See:
http://press.web.cern.ch/Press/Releases00/PR08.00ELEPRundelay.html

\bibitem{feynman}  R. P. Feynman, R. B. Leighton and M. Sands, The Feynman
Lectures on Physics, Vol. 2, Addison-Wesley, New York, 1964, p.
28-4.

\bibitem{thomson}  J. J. Thomson, Phil. Mag., 11, 229 (1881).

\bibitem{heaviside}  O. Heaviside, The Electrician, 14, 220 (1885).

\bibitem{searle}  G. F. C. Searle, Phil. Mag., 44, 329 (1897).

\bibitem{lorentz}  H. A. Lorentz, Proceedings of the Academy of Sciences of
Amsterdam, 6, 809 (1904); Theory of Electrons, 2nd ed. (Dover, New York,
1952).

\bibitem{poincare}  H. Poincar\'{e}, Compt. Rend., 140, 1504 (1905);
Rendiconti del Circolo Matematico di Palermo 21, 129 (1906).

\bibitem{abraham}  M. Abraham, The Classical Theory of Electricity and
Magnetism, 2nd ed. (Blackie, London, 1950).

\bibitem{fermi}  E. Fermi, Nuovo Cimento, 22, 176 (1921); Phys. Zeits., 23,
340 (1922); Rend. Acc. Lincei (5), 31, 184; 306 (1922); Nuovo Cimento, 25,
159 (1923).

\bibitem{mandel}  H. Mandel, Z. Physik, 39, 40 (1926).

\bibitem{wilson}  W. Wilson, Proc. Phys. Soc., 48, 736 (1936).

\bibitem{pryce}  M. H. L. Pryce, Proc. Roy. Soc., A168, 389 (1938).

\bibitem{kwal}  B. Kwal, J. Phys. Rad., 10, 103 (1949).

\bibitem{rohrlich}  F. Rohrlich, Am. J. Phys., 28, 639 (1960); Classical
Charged Particles, (Addison-Wesley, New York, 1990).

\bibitem{history}  On the historical development of the classical
electromagnetic mass theory see \cite{rohrlich} and \cite{augustine}.

\bibitem{butler}  J. W. Butler, Am. J. Phys. 37, 1258 (1969).

\bibitem{poincare2}  In order to account for the stability of the classical
electron Poincar\'{e} \cite{poincare} assumed that part of the electron mass
(regarded as mechanical) originated from forces (known as the Poincar\'{e}
stresses) holding the electron charge together and that it was this
mechanical mass that compensated the $4/3$ factor (reducing the electron
mass from $4/3m$ to $m$). However, the $4/3$ factor, as discussed above,
turned out to be an error in the calculations of electromagnetic mass as
shown in \cite{fermi}-\cite{rohrlich}. As there remained nothing to be
compensated (in terms of mass), if there were some unknown attraction forces
(the Poincar\'{e} stresses) responsible for holding the electron charge
together, their negative contribution to the electron mass would result in
reducing it from m to $2/3m$. This made the stability problem even more
puzzling - on the one hand, a spherical electron tends to disintegrate due
the repulsion of the different parts of the spherical shell; on the other
hand, however, an assumption that there is a force that prevents the
electron charge from blowing up leads to a wrong expression for its mass.
Obviously, there is an implicit assumption in the classical model of the
electron that leads to such a paradox - it is assumed that at every instant
the electron charge occupies the whole spherical shell (see \cite{elementary}%
).

\bibitem{griffiths}  D. J. Griffiths, Introduction to Electrodynamics, 2nd
ed., Prentice Hall, New Jersey, 1989, p. 439.

\bibitem{elementary}  It is not impossible for an $elementary$ charge to
have a spherical but not continuous distribution. Such a possibility follows
from a work \cite{nasko} which has received little attention so far. By
bringing the idea of atomism to its logical completion (discreteness not
only in space but in time as well - $4-$atomism), it is argued in that work
that a quantum-mechanical description of the electron itself (not only of
its state) is possible if the electron is represented not by its worldline
(as deterministically described in special relativity) but by a set of
four-dimensional points (modeled by the energy-momentum tensor of dust - in
this case a sum of delta functions) scattered all over the spacetime region
in which the wave function of the electron is different from zero. The $4-$%
atomism hypothesis gives an insight into two questions: (i) how an
elementary charge can have ''parts'' and still remain an elementary charge,
and (ii) why there is no stability problem despite that the ''parts'' of an
electron repel one another: since for 1 second an electron is represented by
$10^{20}$ four-dimensional points (according to the $4-$atomism hypothesis)
at one instant the electron exists as a single point carrying a greater
(bare) charge, but for one second, for example, there will be $10^{20}$ such
points occupying a spherical shell that manifest themselves as an electron
whose effective charge is equal to the elementary charge. The $4-$atomistic
model of the electron appears to overcome the difficulties of both a purely
particle and a purely wave models of the quantum object and may be a
candidate for what Einstein termed ''something third'' (neither a particle
nor a wave). For a brief description of why neither the purely particle nor
the purely wave models of the quantum object can be accepted see reference
\cite{madelung}.

\bibitem{nasko}  A. H. Anastassov, ''The Theory of Relativity and the
Quantum of Action ($4-$Atomism)'', Doctoral Thesis, Sofia University, 1984
(unpublished); ''Self-Contained Phase-Space Formulation of Quantum Mechanics
as Statistics of Virtual Particles'', Annuaire de l'Universite de Sofia
''St. Kliment Ohridski'', Faculte de Physique, 81, 1993, pp. 135-163.

\bibitem{madelung}  Here are two of the most serious problems with a purely
particle and purely wave models of the electron. (i) If the $s$-electron in
the hydrogen atom is regarded as a particle, i.e. as localized (its charge
being localized) somewhere above the proton, then the hydrogen atom should
possess a dipole moment in its $s$-state. Both quantum mechanics and the
experiment show that this is not the case. One may picture the electron in
the $s$-state as so rapidly orbiting the proton that what is experimentally
measured is the average value of the dipole moment over the measurement
time. And since there is a spherical symmetry in the $s$-state all dipole
moments cancel out exactly - the average value is zero. To verify that
hypothesis Madelung calculated the orbital velocity of the electron that
would ensure that all dipole moments during the measurement cancel out. It
turned out that the electron orbital velocity should be several orders of
magnitude greater than the velocity of light. This shows that the electron
charge should be somehow uniformly distributed around the proton. (ii) A
system of $n$ ''particles'' cannot be represented by a pure wave since that
wave cannot be a real wave in the real space - it is a wave in a space of $3n
$ dimensions.

\bibitem{feynman1} The Feynman Lectures on Physics is not the only physics textbook that stresses
that experimental fact; see, for example, R. Stevenson and R. B. Moore,
Theory of Physics, W. B. Saunders, Philadelphia and London, 1967 (p. 590:
''there is experimental evidence for the existence of electromagnetic
mass'').

\bibitem{moylan}  E. Fermi, Z. Physik, 23, 340-346 (1922), quoted by P.
Moylan, Amer. J. Phys., 63, 818 (1995).

\bibitem{pearle}  P. Pearle, in D. Teplitz, ed. Electromagnetism: Paths to
Research, Plenum Press, New York, 1982, p. 213.

\bibitem{mach}  E. Mach, Science of Mechanics, 9th ed., Open Court, London,
1933. Around 1883 Mach argued that inertia was caused by all the matter in
the Universe (no matter how distant it may be) thus assuming that inertia
had a non-local cause.

\bibitem{f=ma}  The self-force ${\bf F}_{self}$ $=-m{\bf a}$ is
traditionally called inertial force. According to Newton's third law the
external force ${\bf F}$ that accelerates the electron and the self-force $%
{\bf F}_{self}$ have equal magnitudes and opposite directions: ${\bf F}=-$ $%
{\bf F}_{self}$. Therefore ${\bf F}=m{\bf a}$ which means that Newton's
second law is derived on the basis of Maxwell's electrodynamics and Newton's
third law.

\bibitem{repulsion}  Strictly speaking, in the framework of classical
electrodynamics the explanation of the self-force acting on an accelerating
charge appears possible only in terms of repulsion of charges, not in terms
of interaction of a charge and a distorted electric field.

\bibitem{phd}  V. Petkov, Ph. D. Thesis, Concordia University, Montreal,
1997; for an account of why all arguments against regarding the entire mass
of the classical electron as electromagnetic in origin have been answered
see also: ''Acceleration-dependent electromagnetic self-interaction effects
as a basis for inertia and
gravitation''(http://xxx.lanl.gov/abs/physics/9909019)

\bibitem{petkov}  V. Petkov, What is general relativity silent on?
(http://xxx.lanl.gov/abs/gr-qc/0005084).

\bibitem{synge}  J. L. Synge, Relativity: the general theory, Nord-Holand,
Amsterdam, 1960, Ch. III. Sec. 3.

\bibitem{graven}  The formalism of general relativity refuses to yield an
appropriate mathematical (tensor) expression for the energy and
momentum of gravitational field; instead a pseudo-tensor to model
gravitational energy and momentum is used. The problem with a
pseudo-tensor is that it cannot represent a real physical
quantity; this implies that there is no gravitational energy and
momentum. Such a conclusion appears to be fully in line with the
way general relativity describes gravitation - as a manifestation
of spacetime curvature, not as a force field which possesses
energy and momentum; gravitational filed in general relativity
can be regarded as an energy-less and momentum-less geometric
field. However, some people think that there is a real problem
with such a conclusion since the experimental evidence seems to
demonstrate the existence of gravitational energy and momentum -
it is sufficient to mention only the tidal electric power
stations converting what appears to be gravitational energy into
electric energy. That experimental evidence looks completely
differently from the viewpoint of the electromagnetic mass theory
which reveals that gravitation is electromagnetic in origin (at
least in part) - the tidal electric power stations convert
electric energy into electric energy (as in the cases in which
mechanical energy is converted into mechanical energy). There are
attempts to explain the fact that a pseudo-tensor is used for
modeling gravitational energy and momentum: "At issue is not the
existence of gravitational energy, but the localizability of
gravitational energy. It is not localizable" \cite{mtw}. Howeve,
those attempts appear to be a result of a confusion because it is
essentially claimed that two mutually excluding things are true:
(i) there is gravitational energy, and (ii) there is no (force)
gravitational field which means that there is no gravitational
energy.

\bibitem{mtw}  C. W. Misner, K. S. Thorne and J. A. Wheeler, Gravitation,
Freeman, San Francisco, 1973, p.467.

\bibitem{distortion}  An inertial observer with respect to whom an electron
moves with constant speed will see its field deformed (Lorentz-contracted),
but this does not mean that the electron resists its motion with uniform
velocity. The observed deformation of the electron field gives rise to the
increase of the electron mass as seen by the inertial observer.

\bibitem{hrp}  B. Haisch, A. Rueda and H. E. Puthoff, Phys. Rev. A 49, 678
(1994); A. Rueda and B. Haisch, Phys. Lett. A 240, 115 (1998); A. Rueda and
B. Haisch, Found. Phys. 28, 1057 (1998).

\bibitem{stephani}  H. Stephani, General Relativity, 2nd ed., (Cambridge
University Press, Cambridge, New York, 1990), p. 69.
\end{thebibliography}
\end{document}